\documentclass[sigconf]{acmart}
\usepackage{graphicx}
\usepackage{algorithm}
\usepackage{rotating}
\usepackage{balance}

\usepackage{tabularx}
\usepackage{subfigure}
\usepackage{bm}
\usepackage{multirow}
\usepackage{enumitem}
\AtBeginDocument{%
  }

\copyrightyear{2025}
\acmYear{2025}
\setcopyright{acmlicensed}\acmConference[CIKM '25]{Proceedings of the 34th ACM International Conference on Information and Knowledge Management}{November 10--14, 2025}{Seoul, Republic of Korea}
\acmBooktitle{Proceedings of the 34th ACM International Conference on Information and Knowledge Management (CIKM '25), November 10--14, 2025, Seoul, Republic of Korea}
\acmDOI{10.1145/3746252.3760833}
\acmISBN{979-8-4007-2040-6/2025/11}

\begin{document}

\title{Asymmetric Diffusion Recommendation Model}

\author{Yongchun Zhu}
\affiliation{%
  \institution{ByteDance}
  \city{Beijing}
  \country{China}}
\email{zhuyongchun.zyc@bytedance.com}

\author{Guanyu Jiang}
\affiliation{%
  \institution{ByteDance}
  \city{Beijing}
  \country{China}}
\email{jiangguanyu.jgy@bytedance.com}

\author{Jingwu Chen$*$}
\affiliation{%
  \institution{ByteDance}
  \city{Beijing}
  \country{China}}
\email{chenjingwu@bytedance.com}

\author{Feng Zhang}
\affiliation{%
  \institution{ByteDance}
  \city{Shanghai}
  \country{China}}
\email{feng.zhang@bytedance.com}

\author{Xiao Yang}
\affiliation{%
  \institution{ByteDance}
  \city{Beijing}
  \country{China}}
\email{wuqi.shaw@bytedance.com}

\author{Zuotao Liu}
\affiliation{%
  \institution{ByteDance}
  \city{Shanghai}
  \country{China}}
\email{michael.liu@bytedance.com}

\thanks{$*$ Jingwu Chen is the corresponding author.}

\renewcommand{\shortauthors}{Yongchun Zhu et al.}

\begin{abstract}
  Recently, motivated by the outstanding achievements of diffusion models, the diffusion process has been employed to strengthen representation learning in recommendation systems. Most diffusion-based recommendation models typically utilize \textit{standard Gaussian noise} in \textit{symmetric} forward and reverse processes in continuous data space. Nevertheless, the samples derived from recommendation systems inhabit a discrete data space, which is fundamentally different from the continuous one. Moreover, Gaussian noise has the potential to corrupt personalized information within latent representations. In this work, we propose a novel and effective method, named Asymmetric Diffusion Recommendation Model (\textbf{AsymDiffRec}), which learns forward and reverse processes in an asymmetric manner. We define a generalized forward process that simulates the missing features in real-world recommendation samples. The reverse process is then performed in an asymmetric latent feature space. To preserve personalized information within the latent representation, a task-oriented optimization strategy is introduced. In the serving stage, the raw sample with missing features is regarded as a noisy input to generate a denoising and robust representation for the final prediction. By equipping base models with AsymDiffRec, we conduct online A/B tests, achieving improvements of +0.131\% and +0.166\% in terms of users' active days and app usage duration respectively. Additionally, the extended offline experiments also demonstrate improvements. AsymDiffRec has been implemented in the Douyin Music App.
\end{abstract}

\ccsdesc[500]{Information systems~Recommendation systems}

\keywords{Recommendation, Diffusion Model}

\maketitle

\section{Introduction}\label{sec:1}

Recently, deep neural-network models have surged to the forefront of modern recommendation systems, permeating both the industry and academia, and improving deep representation learning has emerged as a pivotal topic in recommendation systems. To enhance representation learning, Autoencoder-based methods~\cite{wu2016collaborative,zhuang2017representation,liang2018variational,xu2022alleviating} which recover the original representation from latent vectors have been widely utilized. In addition, self-supervised learning which enhances representations from data without the need for explicit human-annotated labels has demonstrated its great capacity for representation learning in recommendation systems~\cite{zhou2020s3,yao2021self}.

\begin{figure*}[t]
	\centering
	\begin{minipage}[b]{0.92\linewidth}
		\centering
		\includegraphics[width=1.\linewidth]{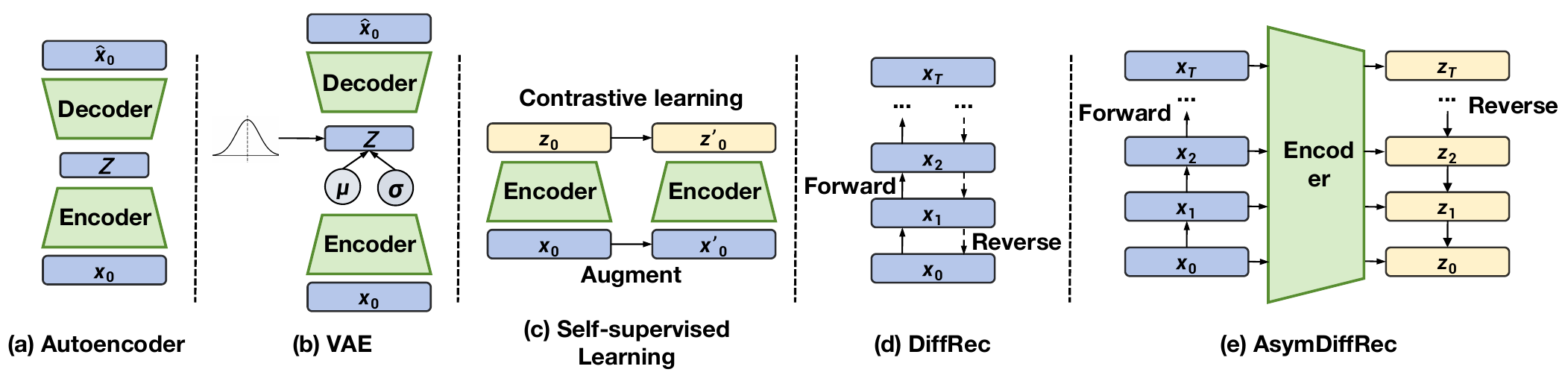}
	\end{minipage}
	\caption{Illustration of (a) Autoencoder, (b) Variational Autoencoders, (c) Self-supervised Method, (d) Diffusion Recommendation (DiffRec), and (e) the proposed AsymDiffRec.}\label{fig:model}
\end{figure*}

Inspired by remarkable achievements of diffusion models in computer vision, some attempts have been undertaken to apply the diffusion process for the purpose of enhancing recommendation models~\cite{wang2023diffusion,zhao2024denoising,jiang2024diffkg}. However, these methods typically utilize \textit{standard Gaussian noise} within \textit{symmetric} forward and reverse processes, which could limit the capabilities of representation learning in recommendation systems. Moreover, it is observed the performance of diffusion models with Gaussian noise is still deficient in real-world large-scale industrial datasets. We argue that there are two underlying reasons:
\begin{itemize}[leftmargin=1em]
    \item \textbf{Discrete data space}: Existing Diffusion methods using standard continuous Gaussian noise show satisfactory performance in continuous data space, e.g., image signals. Nevertheless, the data generated by recommendation systems resides in discrete space, e.g., gender, ID, and sequence. Thus, the augmented representation, generated by adding continuous Gaussian noise to the latent representation extracted from a real sample, is incapable of representing another real-world sample. Representations that are robust to Gaussian noise have limited effectiveness for recommendation systems.
    \item \textbf{Corrupt personalized information}: These methods add and reconstruct Gaussian noise within symmetric forward and reverse processes, which may cause the model to focus on Gaussian noise and overlook personalized information. However, learning personalized information is of utmost significance for a personalized online recommendation service. 
\end{itemize}

To solve the above issues, we propose an Asymmetric Diffusion Recommendation Model (AsymDiffRec). The training stage of AsymDiffRec contains a forward process and a reverse process, similar to most diffusion methods. To adapt the diffusion process to the discrete data space of recommendation systems, AsymDiffRec uses a discrete operation to add noise to the raw data in each step of the forward process. We define the discrete operation as feature dropout. In each step, one feature is dropped, which is capable of simulating the scenario where features are missing in real samples. Then, the reverse process is performed in an asymmetric way to reconstruct the latent representation of the original data from the noisy latent representation, which can be regarded as a process of `feature completion'. In addition, to preserve personalized information, AsymDiffRec utilizes an auxiliary task-oriented loss based on the reconstructed latent representation.


Note that most existing diffusion-based recommendation methods employ the diffusion module for auxiliary training to learn robust representations. In contrast, AsymDiffRec utilizes the diffusion module not only in the training stage but also in the serving stage.
In practical recommendation systems, there is the problem of missing features in most samples. Thus, in the serving stage of AsymDiffRec, the input sample is regarded as a noisy sample with $T$ missing features, and AsymDiffRec reconstructs a complete representation through the reverse process for the final prediction. The comparison between AsymDiffRec and existing methods is shown in Figure~\ref{fig:model}. The main contributions of our work are summarized in two folds:
\begin{itemize}[leftmargin=1em]
    \item To adapt the diffusion model to industrial recommendation systems, we propose a new method named Asymmetric Diffusion Recommendation Model (AsymDiffRec). 
    \item We conduct online experiments, obtaining +0.131\% and +0.166\% in terms of users' active days and app usage duration respectively. In addition, offline experiments also demonstrate its effectiveness. AsymDiffRec has been deployed in online recommendation systems of Douyin Music App.
\end{itemize}

\section{Related Work}

In recommendation systems, representation learning plays a crucial role, which aims to learn effective and personalized representations of users and items. Improving deep representation learning has attracted a great deal of attention. Traditional methods~\cite{wu2016collaborative,zhuang2017representation} enhance representation learning with Auto-Encoders, which train a neural network to reconstruct a data point from the latent representation. ~\cite{liang2018variational,xu2022alleviating,xie2021adversarial} extend Variational Auto-Encoders to collaborative filtering, which mainly learn an encoder for posterior estimation, and a decoder to reconstruct original data. Self-supervised learning shows great success in representation learning, which has been widely used in recommendation systems~\cite{zhou2020s3,yao2021self,guo2022miss}. ~\citet{zhu2025afm} proposed a model-agnostic framework AdaF$^2$M$^2$, which achieves comprehensive feature learning via multi-forward training with augmented samples, while the adapter applies adaptive weights on features responsive to different user/item states.

Recently, diffusion models have achieved remarkable performance in the field of computer vision, and some researchers make efforts to adapt diffusion models to recommendation systems~\cite{wang2023diffusion,liu2023diffusion,zhao2024denoising}. However, most of these methods add and reconstruct Gaussian noise, which could corrupt personalized information within the discrete data space of recommendation systems. ~\citet{lin2024discrete} proposed a discrete diffusion for the rerank task, and ~\citet{wu2019neural} consider information diffusion on social networks. Different from these works~\cite{lin2024discrete,wu2019neural}, we focus on learning robust representations for universal ranking recommendation models. To the best of our knowledge, we are the first to use diffusion-based models to achieve improvement in active days in industry.

\section{Asymmetric Diffusion Recommendation Model}

In this section, we present details of the proposed Asymmetric Diffusion Recommendation Model (AsymDiffRec). AsymDiffRec incorporates a forward process, during which discrete noise is added within the raw feature space, and an asymmetric reverse process in a latent space.

\subsection{Problem Statement}\label{sec:3.1}
In a standard binary classification task in recommendation systems, each sample contains a label $y \in \{0,1\}$ and the input raw features $\bm{x} = \{x_1, \cdots, x_N\}$, where $N$ indicates the number of raw features. We define a feature extractor which contains embedding layers and deep networks as $h(\cdot)$. The extracted latent representation is denoted as $\bm{z} = h(\bm{x})$. The final prediction is performed based on the latent vector $\hat{y} = f(\bm{z})$. The cross-entropy loss is often used as an optimization objective for binary classification:
\begin{equation}
    \mathcal{L}_{main} = -y \log \hat{y} - (1-y) \log (1 - \hat{y}).\label{eq:loss}
\end{equation}

\begin{table*}[htbp]
\centering
\setlength\tabcolsep{3pt}
\caption{Online A/B testing results of a ranking task. The results indicate the relative improvement with AsymDiffRec over the baseline (a DCN-V2-based multi-task model). The square brackets represent the 95\% confidence intervals for online metrics. Statistically significant improvement is marked with bold font in the table.}
\begin{tabular}{lcccccc}
\toprule
\multirow{2}{*}{}                    & \multicolumn{2}{c}{Main Metrics}                    & \multicolumn{4}{c}{Constraint Metrics}                                                                    \\
\cmidrule(r){2-3}  \cmidrule(r){4-7}
                                     & Active Day               & Duration                 & Like                     & Finish                   & DisLike                 & Play                     \\
\midrule
\multirow{2}{*}{AsymDiffRec}             & \textbf{0.131\%}         & \textbf{0.166\%}         & \textbf{0.285\%}         & \textbf{0.353\%}         & -2.469\%         & \textbf{0.256\%}         \\
                                     & {[}-0.041\%, +0.041\%{]} & {[}-0.092\%, +0.092\%{]} & {[}-0.116\%, +0.116\%{]} & {[}-0.104\%, +0.104\%{]} & {[}-3.001\%, +3.001\%{]} & {[}-0.095\%, +0.095\%{]}\\
\bottomrule
\end{tabular}\label{tab:online}
\end{table*}

\begin{table}[htbp]
\centering
\caption{Offline results (AUC, UAUC and RelaImpr) on the industrial datasets DouyinMusic-40B.}
\begin{tabular}{lcccc}
\toprule
               & AUC  & RelaImpr  & UAUC  & RelaImpr   \\
\midrule
Base Model    & 0.92267 & - &  0.61578 & - \\
CDAE    & 0.92220 & -0.051\% & 0.61600 & +0.036\% \\
MultiVAE & 0.92285 & +0.020\% &  0.61834 & +0.416\% \\
SSL & 0.92297 & +0.033\% & 0.62009 & +0.700\%\\
DiffRec & 0.92307 & +0.043\% & 0.62152 & +0.932\%\\
\midrule
AsymDiffRec & \textbf{0.92359} & \textbf{+0.100\%} & \textbf{0.62614} & \textbf{+1.682\%} \\
w/o recon & 0.92276 & +0.010\% & 0.62025 & +0.726\% \\
w/o aux & 0.92248 & -0.021\% & 0.62069 & +0.797\% \\
\bottomrule
\end{tabular}\label{tab:offline}
\end{table}

\subsection{Discrete Forward Process}\label{sec:3.2}

The forward process of most diffusion-based recommendation models adds standard Gaussian noise similar to diffusion models in computer vision. However, it is sub-optimal for recommendation systems due to the discrete characteristics. Thus, we propose a discrete forward process for universal ranking models.

We use the input raw data as the original input $\bm{x}_0$. With iterative $T$ forward steps, the noisy data can be denoted as $\{\bm{x}_1, \cdots, \bm{x}_T\}$, where $T$ is randomly sampled in $\text{Uniform}(0, N)$. For diffusion models, it is important to define one step of the forward process $q(\bm{x}_t|\bm{x}_{t-1})$. To enable the forward process to simulate real-world noise, we develop a discrete operation at the feature level. Note that the problem of missing features is common in practical recommendation systems. Therefore, we define a discrete operation as \textbf{feature dropout} that drops a feature with a uniform probability in each forward step. In other words, $T$ forward steps mean that $T$ features are dropped from $\bm{x}_0$ and the noisy data $\bm{x}_T$ is obtained. The noisy sample $\bm{x}_T$ has the potential to simulate a real sample in recommendation systems, and this approach holds greater significance compared to simply adding standard Gaussian noise to $\bm{x}_0$.

\subsection{Asymmetric Reverse Process}\label{sec:3.3}

The reverse process in diffusion models aims to reconstruct the original sample $\bm{x}^{\prime}_0$ from the noisy sample $\bm{x}_T$ with a denoising function $g(\cdot)$. Existing methods perform the forward and reverse processes in a symmetric manner and within the same data space. However, the recommendation model makes predictions based on latent representations, and it is important to extract latent representations from the reconstructed sample $\bm{x}^{\prime}_0$. To mitigate information loss that occurs in two stages, including a reverse process in the raw data space and a feature extraction process, we propose a asymmetric reverse process. The reverse process works directly in the latent representation space, which is different from the forward process that operates in the raw feature space.

To facilitate the learning of the denoising function, we define a step embedding $\bm{s} = [0,1,1,\cdots,0,1]$, where $1$ means that the corresponding feature is missing. The denoising function takes the step embedding $\bm{s}$ and the noisy representation $\bm{z}_T = h(\bm{x}_T)$ as input to generate the denoising representation $\bm{z}_0^{\prime} = g([\bm{s},\bm{z}_T])$. Then, the reconstruction loss is formulated as:
\begin{equation}
    \mathcal{L}_{recon} = || \bm{z}_0^{\prime} - \bm{z}_0 ||^2, \label{eq:reconstruct_loss}
\end{equation}
where $\bm{z}_0$ denotes the latent representation of the original sample, and $\bm{z}_0^{\prime}$ indicates the latent representation of the noisy sample. Thus, by minimizing Equation~(\ref{eq:reconstruct_loss}), the denoising function is enabled to learn to reconstruct representations of samples with all features from noisy representations of samples that have $T$ missing features. Note that the step embedding $\bm{s}$ can provide the position information of missing features, which is beneficial for model learning process.

To preserve personalized information in the reconstructed representation, we introduce an auxiliary task-oriented loss to guide the learning of the denoising function:
\begin{equation}
    \mathcal{L}_{aux} = -y \log f(\bm{z}_0^{\prime}) - (1-y) \log (1 - f(\bm{z}_0^{\prime})).\label{eq:aux_loss}
\end{equation}

\begin{algorithm} [t] 
    \caption{Asymmetric Diffusion Recommendation Model}\label{alg}
    \flushleft{
    \hspace*{0.02in}\textbf{Training Stage}:\\
    \hspace*{0.04in} \textbf{repeat} \\
    \hspace*{0.10in} 1.Sample $T$ from a uniform distribution $\text{Uniform}(0, N)$;\\
    \hspace*{0.10in} 2.Compute the noisy representation $\bm{z}_T$ via discrete forward process;\\
    \hspace*{0.10in} 3.Compute $\mathcal{L}_{recon}$ and $\mathcal{L}_{aux}$ via asymmetric reverse process;\\
    \hspace*{0.10in} 4.Learn the overall framework with  $\mathcal{L}_{main} + \mathcal{L}_{recon} + \mathcal{L}_{aux}$;\\
    \hspace*{0.04in} \textbf{until} converged\\
    \hspace*{0.02in}\textbf{Serving Stage}: \\
    \hspace*{0.10in} 5.Compute $\bm{z}_0^{\prime} = g([\bm{s}, h(\bm{x}_0)])$ for the final prediction;\\
    }
\end{algorithm}

\subsection{Overall Framework}
The deployment of the AsymDiffRec framework consists of two stages: the training stage and the serving stage. The overall training and serving procedure is summarized in Algorithm~\ref{alg}. In the training stage, we train all parameters in AsymDiffRec by minimizing $\mathcal{L}_{main} + \mathcal{L}_{recon} + \mathcal{L}_{aux}$.

Most existing diffusion-based recommendation methods utilize the diffusion module as an auxiliary learning task to enhance the latent representations. Different from existing methods, AsymDiffRec makes use of the diffusion module not only during the training stage but also in the serving stage. The problem with input features for online serving is usually that features are missing. Thus, in the online serving stage, the input features $\bm{x}_0$ are treated as the noisy one, and the denoising representation can be calculated by $\bm{z}_0^{\prime} = g([\bm{s}, h(\bm{x}_0)])$, which is then used for the final prediction. It should be noted that the denoising function can be regarded as a process of feature completion. In our online system, the dimension of $\bm{z}$ is $128$, and the denoising function is a simple two-layer neural network. Thus, in comparison to the online baseline, the inference time of the proposed diffusion model shows only a marginal increase (average inference time 12.6 ms v.s. 12.1 ms).

\section{Experiments}

In this section, we conduct extensive offline and online experiments with the aim of answering the following evaluation questions: 
\begin{itemize}
     \item[\textbf{EQ1}] Can AsymDiffRec bring improvements to the performance of online recommendation tasks?
     \item[\textbf{EQ2}] Does AsymDiffRec outperform other approaches in offline datasets?
     \item[\textbf{EQ3}] What are the effects of the reconstruction loss and the auxiliary task-oriented loss in our proposed AsymDiffRec?
\end{itemize}

\textbf{Datasets.} We evaluate AsymDiffRec with baselines on a large industrial recommendation dataset. Note that AsymDiffRec requires a large amount of features for the forward process with the feature dropout mechanism. However, there is currently no public dataset that is suitable for this purpose. Thus, we only adopt the industrial dataset DouyinMusic-40B from our system.

\textit{DouyinMusic-40B}: Douyin offers a music recommendation service, which has over 10 million daily active users. We collect data from the impression logs and get one dataset. The dataset contains more than 40 billion samples, denoted as \textit{DouyinMusic-40B}. Each sample in the industrial dataset includes over two hundred features. These features consist of non-ID meta features such as gender, age, genre, mood, scene, and various ID-based personalized features. The DouyinMusic-40B dataset contains samples from Douyin Music over a time span of 10 weeks from September to October 2024. We use `Collection' as the label. 

\textbf{Online A/B Testing (EQ1).} 
Following~\cite{zhu2024interest,zhu2025long,zhu2025afm}, we evaluate model performance based on two main metrics, Active Days and Duration, which are widely adopted in practical recommendation systems. The total days that users in the experimental bucket open the application are denoted as Active Days. The total amount of time spent by users in the experimental bucket on staying in the application is denoted as Duration. We also take additional metrics, which evaluate user engagement, including Like/Dislike (clicking the like/dislike button on the screen), Finish (hearing the end of a song), and Play (play a song), which are usually used as constraint metrics. We calculate all online metrics per user.

To measure the improvements that AsymDiffRec brings to our online music service, we conducted comprehensive one-month A/B testing with 100\% of users (more than 20 million) for the ranking task in Douyin Music App. We apply the proposed AsymDiffRec on a DCN-V2-based model~\cite{wang2021dcn} which is deployed in the online ranking tasks. The online A/B results are shown in Table~\ref{tab:online}. For the main metrics Active Days and Duration, the proposed AsymDiffRec achieves a large improvement of +0.131\% and +0.166\% for all users with statistical significance. AsymDiffRec has also demonstrated significant improvements in constraint metrics, such as increased Like, Finish, and Play, which indicates that AsymDiffRec can enhance user engagement and satisfaction.

\textbf{Offline Results (EQ2).} 
For binary classification tasks, AUC is a widely used metric~\cite{fawcett2006introduction}. It measures the goodness of order by ranking all the items with prediction, including intra-user and inter-user orders. Besides, AUC is a common metric for recommendation~\cite{he2017neural,zhou2018deep}. In addition, we introduce another metric in our business, named UAUC, which calculates AUC per user and then averages scores with weights proportional to the user's sample size. A larger UAUC suggests better model performance. For the industrial datasets, we report the relative improvement (RelaImpr) of AUC and UAUC over base models.

We compare the proposed AsymDiffRec with (1) Base Model which is a DCN-V2-based model~\cite{wang2021dcn}, (2) CDAE~\cite{wu2016collaborative}, (3) MultiVAE~\cite{liang2018variational}, (4) Self-supervised learning (SSL)~\cite{yao2021self}, and (5) DiffRec~\cite{wang2023diffusion}. The experimental results on the industrial dataset are shown in Table~\ref{tab:offline}. AsymDiffRec significantly outperforms CDAE, MultiVAE, and DiffRec. The improvements come from that AsymDiffRec can learn more useful knowledge from real augmented samples. SSL~\cite{yao2021self} also generates real augmented samples. However, AsymDiffRec attains superior results, which indicates AsymDiffRec is more capable of learning robust representations.

\textbf{Ablation Study(EQ3)}. To test the effectiveness of each loss in AsymDiffRec, we present an ablation study. We introduce two kinds of models, AsymDiffRec w/o recon and w/o aux that indicate removing the reconstruction loss and the auxiliary task-oriented loss.
The results show that AsymDiffRec achieves better performance than AsymDiffRec w/o recon and w/o aux, which demonstrates each loss
is effective. Note that AUC of AsymDiffRec w/o aux performs worse than that of the Base Model, which demonstrates preserving personalized information is important for enhancing representation learning in recommendation systems.

\section{Conclusion}
In this paper, for enhancing representation learning in recommendation systems, we propose the Asymmetric Diffusion Recommendation Model (AsymDiffRec), a novel and effective diffusion-based method that learns forward and reverse processes in an asymmetric manner.
Different from existing methods using Gaussian noise, we define a discrete noise as feature dropout in raw data space. In addition, the reverse process is operated in another latent feature space. In the serving stage, the raw sample with
missing features is regarded as a noisy input to generate a denoising
and robust representation for the final prediction.
We demonstrated the superior performance of the proposed AsymDiffRec in both offline and online experiments. The online experiments achieve 0.131\% and 0.166\% improvements on users' active days and app duration respectively, which demonstrates the effectiveness of AsymDiffRec in online systems. Moreover, AsymDiffRec has been deployed in the Douyin Music App.

\bibliographystyle{ACM-Reference-Format}
\bibliography{sample-base}

\end{document}